\documentclass[onecolumn,10pt]{IEEEtran}
\usepackage[normalem]{ulem}
\usepackage{multirow}
\usepackage[cmex10]{amsmath}
\usepackage{amssymb}

\providecommand{\remarkname}{\bf{Remark}}
\usepackage{graphicx}
\usepackage{tikz}

\usepackage{algorithm,algorithmic}
\usepackage{tabularx}
\usepackage[skip=2pt,font=small]{caption}
\usepackage{color,soul}
\usepackage{caption}
\usepackage[position=top]{subfig}
\usepackage{subfig,float}
\usepackage{epstopdf}
\usepackage{diagbox}

\usepackage{multirow}
\usepackage{multicol}

\usepackage{mathtools}
\usepackage{scalerel}

\begin{document}
\title{Supplementary Notes: Segment Parameter Labelling in MCMC Change Detection}
\author{Alireza Ahrabian
}
\maketitle

\begin{abstract}
\indent This work addresses the problem of segmentation in  time series data with respect to a statistical parameter of interest in Bayesian models. It is common to assume that the  parameters are distinct  within each segment. As such, many Bayesian change point detection models do not exploit the segment parameter patterns, which can improve performance. This work proposes a Bayesian  change point detection algorithm that makes use of repetition in  segment parameters, by introducing segment class labels that utilise a Dirichlet process prior. 
\end{abstract}
\section{Introduction}
Change  detection algorithms are an  important tool in the exploratory analysis of real world data. Namely, such algorithms seek to partition  time series data with respect to a statistical parameter of interest, where examples include, the mean, variance and autoregressive model weights, to name but a few.  Change point detection algorithms have found applications in fields ranging from financial to bio-medical  data analysis. 
\\
\indent Accordingly, many  approaches have been proposed to address  the problem of segmenting time series data. Namely, the work in \cite{Brandt83} proposed an online method, based on the likelihood ratio test that both detects and estimates a change in the statistical parameter of interest. More recently, the work in \cite{Killick12} proposed a computationally efficient algorithm that segments  data  by minimising a cost function using dynamic programming, while \cite{Desobry05} utilises a classification algorithm (kernel-SVM) in order to estimate the change point locations. Bayesian approaches to  time series segmentation  have also proven to be useful. In particular, by deriving the posterior distribution of the parameters interest (that includes, the number of segments, as well as, the change point locations), one can then use suitable methodologies (for evaluating the posterior distribution) in order to both infer as well as predict. Examples of such work includes,  a fully hierarchical  Bayesian model proposed in \cite{Punskaya02}, that utilised  a  Markov Chain Monte Carlo (MCMC) sampler in order to evaluate the posterior distribution of the target parameters. While an exact (that is, avoiding the use MCMC sampling algorithms) segmentation algorithm was developed  in \cite{Fearnhead05}, by evaluating the posterior distribution using an recursive algorithm.
\\
\indent  The change point algorithms mentioned in the previous section, assume that statistical parameters of interest from different segments are distinct and thus  independent. However, many real world processes can often be modeled by parameters generated from a fixed number of states, where such parameters  can be re-assigned more than once (parameter repetition). In particular, hidden Markov models (HMM) and their extensions \cite{Rezek2005},  assign to each data point a state label (that evolves according to Markov chain) corresponding to a set of parameters that govern the emission probability of generating the data point. While, mixture models \cite{Neal00}\cite{Rasmussen00} and their extensions (e.g. Dirichlet processes \cite{Neal00}) assume that each data point is generated from a probability distribution; with the parameters of the distribution belong to state drawn from a discrete distribution (that captures the clustering of the data points, with respect to the parameter of interest). However, it should be noted that such methods  assign a parameter belonging to a particular state to each time point, and not to the parameters corresponding to a given segment.  
\\
\indent This work is based on the method outlined in \cite{Ahrabian18} that incorporates segment parameter repetition in the estimation process of a change point detection algorithm, when estimating changes in autoregressive processes. Namely, the work proposed to extend the Bayesian change point detection algorithm proposed in \cite{Punskaya02}, by incorporating a  parameter  class variable that utilises a Dirichlet process prior for identifying  the number of distinct segment parameters. By including the parameter class variable, segment parameter repetition is captured during the estimation process of the transition times (change point locations), resulting in more robust segmentation. 
\section{Background}
\subsection{MCMC Change Point Detection}
  In this work, we consider the change point detection algorithm proposed in \cite{Punskaya02}. That is, given a set of  transition times $\boldsymbol{\tau}_{K}=[\tau_{1},...,\tau_{K}]$ where $\tau_{0}=1$ and $\tau_{K+1}=N$, that partition a data set  $\mathbf{x}$  into $K+1$ segments, where for each segment (consisting of data points between the time indices $\tau_{i}+1\leq \tau \leq \tau_{i+1}$)  there exists the following functional relationship  between the data points $\boldsymbol{\text{x}}_{\tau_{i}+1:\tau_{i+1}}$ and  the statistical parameter $\boldsymbol{\phi}_{i}\in \mathbb{R}^{D}$, that is 
\begin{equation}
\boldsymbol{\text{x}}_{\tau_{i}+1:\tau_{i+1}}=f_{d}(\boldsymbol{\text{x}}_{\tau_{i}+1:\tau_{i+1}},\boldsymbol{\phi}_{i}) + \boldsymbol{\text{n}}_{\tau_{i}+1:\tau_{i+1}}
\label{model}
\end{equation}
for  segments indexed by $i=\{0,\dots,K\}$, where $\boldsymbol{\text{n}}_{\tau_{i}+1:\tau_{i+1}}$ is a set of i.i.d. Gaussian noise samples with zero mean and variance $\sigma_{i}^{2}$. In particular, an example of the  functional relationship $f_{d}(\boldsymbol{\text{x}}_{\tau_{i}+1:\tau_{i+1}},\boldsymbol{\phi}_{i})$ is given by the autoregressive model of order $D-1$, that is
\begin{equation*}
f_{d}(\boldsymbol{\text{x}}_{\tau_{i}+1:\tau_{i+1}},\boldsymbol{\phi}_{i}) \hspace{-1mm}=\hspace{-1mm}\begin{bmatrix}
    1 & 0 & 0 & \dots  & 0 \\
    1 & x_{\scaleto{\tau_{i}+1}{4pt}} & 0 & \dots  & 0 \\
    \vdots & \vdots & \vdots & \ddots & \vdots \\
    1 & x_{\scaleto{\tau_{i+1}-1}{4pt}} & x_{\scaleto{\tau_{i+1}-2}{4pt}} & \dots  & x_{\scaleto{\tau_{i+1}-p+1}{4pt}}
\end{bmatrix}
\hspace{-2mm}
\begin{bmatrix}
    \phi^{0}_{i}        \\
    \vdots          \\
    \phi^{0}_{D-1}       \\
\end{bmatrix} 
\label{model}
\end{equation*}
where $\boldsymbol{\phi}_{i}=[\phi^{0}_{i} ,...,\phi^{0}_{D-1} ]^{T}$.
Accordingly,  one can define the posterior distribution for the target parameters of interest; namely, the number of transition times $K$ and set of transition time points $\boldsymbol{\tau}_{K}$, that is 
\begin{equation}
p(K,\boldsymbol{\tau}_{K}|\mathbf{x})\propto \int p(\mathbf{x}|\boldsymbol{\Phi},K,\boldsymbol{\tau}_{K})p(\boldsymbol{\Phi},K,\boldsymbol{\tau}_{K}) d\boldsymbol{\Phi}
\label{post_mcmc}
\end{equation}
 where $\boldsymbol{\Phi}=[\boldsymbol{\phi}_{0},...,\boldsymbol{\phi}_{K}]$ corresponds to the vector of segment parameters that is treated as a nuisance parameter and thus integrated out of the posterior distribution. Finally, it should be noted that the  likelihood function  in the posterior distribution  \eqref{post_mcmc}  assumes that the parameters $\boldsymbol{\phi}_{i}$ are distinct for each segment \cite{Punskaya02}, that is  
\begin{equation}
p(\mathbf{x}|\boldsymbol{\Phi},K,\boldsymbol{\tau}_{K})=\prod_{i=0}^{K}p(\mathbf{x}_{\tau_{i}+1:\tau_{i+1}}|\boldsymbol{\phi}_{i})
\label{like_mcmc}
\end{equation}
\subsection{Dirichlet Process Mixture Model}
Consider a set of $N$ exchangeable data points $\mathbf{x}$, such that probability distribution of the data points $p(\mathbf{x})$ can be represented by a set of $V$ class distributions, that is
\begin{equation}
p(\mathbf{x})=\sum_{v=1}^{V}\pi_{v}f(\mathbf{x}|\theta_{v})
\label{MM}
\end{equation}
where $\pi_{v}$ is the mixing coefficient and $\theta_{v}$ corresponds to the class parameter/s of the probability distribution $f(.)$. The Dirichlet process mixture model (DPMM) \cite{Neal00} can be seen as the limiting case of the mixture model (MM) specified in \eqref{MM}.  This can be seen by first considering the re-formulation of the mixture model shown in \eqref{MM}, by introducing  the class indicator random variable $c_{i}$ for the $i^{\text{th}}$ data point
\begin{equation}
\begin{aligned}
x_{i}|c_{i},\boldsymbol{\theta}&\sim f(x_{i}|\theta_{c_{i}})\\
c_{i}|\boldsymbol{\pi}&\sim \text{Discrete}(\pi_{1},\dots,\pi_{V})\\
\theta_{v}&\sim G_{0}\\
 \boldsymbol{\pi}|\alpha &\sim \text{Dir}(\alpha/V,\dots,\alpha/V)
\end{aligned}
\label{gen_model_dirichlet}
\end{equation}
 where  $\boldsymbol{\theta}=[\theta_{1},\dots,\theta_{V}]$, $\boldsymbol{\pi}=[\pi_{1},\dots,\pi_{V}]$ and $G_{0}$ denotes the 
 prior distribution  on the parameters $\theta_{v}$. The probability of selecting a given class $c_{i}$ is determined by the mixing coefficients $\boldsymbol{\pi}$. In particular, the the joint distribution of the class indicator random variables is given by
 \begin{equation}
p(c_{1},...,c_{N}|\boldsymbol{\pi})=\prod_{v=1}^{V}\pi_{v}^{n_{v}}
\label{CRV}
\end{equation}
where $n_{v}$ corresponds to the number of data points assigned to class $v$. The distribution of the the $i^{\text{th}}$ class indicator variable  given all other class variables $\boldsymbol{\text{c}}_{-i}$ that excludes the $c_{i}$, is given by the following
\setcounter{equation}{2}
 \begin{equation}
 \begin{aligned}
p(c_{i}=v|\boldsymbol{\text{c}}_{-i},\alpha)&=\frac{\int p(c_{i}=v,\boldsymbol{\text{c}}_{-i}|\boldsymbol{\pi})p(\boldsymbol{\pi}|\alpha)d\boldsymbol{\pi}}{\int p(\boldsymbol{\text{c}}_{-i}|\boldsymbol{\pi})p(\boldsymbol{\pi}|\alpha)d\boldsymbol{\pi}}\\[5pt]
&=\frac{n_{-i,v}+\alpha/V}{N-1+\alpha}
\end{aligned}
\label{condCRV}
\end{equation}
where $n_{-i,v}$ is the number of data points (excluding $x_{i}$) assigned to class $v$ and $p(\boldsymbol{\pi}|\alpha)$ is a symmetric Dirichlet distribution with parameter $\alpha/V$. 
\\
\indent Taking the number of classes $V\rightarrow  \infty$ and assuming that there exists a  finite number of \textit{represented classes}  $V'$, such that the number of data points assigned to each  class is greater than zero; accordingly the conditional probability in \eqref{condCRV} for represented classes is given by 
 \begin{equation}
p(c_{i}=v|\boldsymbol{\text{c}}_{-i},\alpha)=\frac{n_{-i,v}}{N-1+\alpha}
\label{repClass}
\end{equation}
 That is, \eqref{repClass} is the probability of assigning the class variable $c_{i}$ to the represented class $v$. Furthermore, as $V\rightarrow\infty$  there exists an countably infinite number of classes that excludes the represented classes  such that, $c_{i}\neq c_{l}$, for all $l\neq i$. In order to calculate the probability of assigning $c_{i}$ to a new class, consider the following
 \begin{equation}
 \begin{aligned}
&\sum_{v'=1}^{\infty}p(c_{i}=v'|\boldsymbol{\text{c}}_{-i},\alpha)=1\\
&\sum_{v'=1}^{V'}p(c_{i}=v'|\boldsymbol{\text{c}}_{-i},\alpha)+\sum_{v'=V'+1}^{\infty}p(c_{i}=v'|\boldsymbol{\text{c}}_{-i},\alpha)=1
\end{aligned}
\label{sumCRV}
\end{equation}
where $\sum_{v'=1}^{V'}p(c_{i}=v|\boldsymbol{\text{c}}_{-i},\alpha)$ corresponds to the sum of the probabilities of the represented classes (that is, there exists a data point assigned to the class) and $\sum_{v'=V'+1}^{\infty}p(c_{i}=v|\boldsymbol{\text{c}}_{-i},\alpha)$ the sum of the probabilities of all other classes. Accordingly the probability of assigning $c_{i}$ to a new class,\footnote{The classes that exclude the represented classes.} is given by the following 
 \begin{equation}
 \begin{aligned}
p(c_{i}\neq c_{l} \hspace{1mm} \text{for all}  \hspace{1mm} i\neq l|\boldsymbol{\text{c}}_{-i})&=\sum_{v'=V'+1}^{\infty}p(c_{i}=v|\boldsymbol{\text{c}}_{-i},\alpha)\\
&=1-\sum_{v'=1}^{V'}p(c_{i}=v'|\boldsymbol{\text{c}}_{-i},\alpha)\\
&=\frac{\alpha}{N-1+\alpha}
\end{aligned}
\label{sumCRV1}
\end{equation}
\\
\indent The conditional posterior distribution of the class variable $c_{i}$, given \eqref{repClass} and \eqref{sumCRV1}, can then be determined as follows
\begin{equation}
p(c_{i}=v|\boldsymbol{\text{c}}_{-i},x_{i},\boldsymbol{\theta})\propto\frac{n_{-i,v}}{N-1+\alpha}L(x_{i}|\theta_{v})
\label{cond_post_dirichlet1}
\end{equation}
 for a class where $n_{-i,v}>0$ and corresponds to the likelihood $L(x_{i}|\theta_{v})$. The conditional posterior probability for assigning a data point to a new class is given by
\begin{equation}
\begin{aligned}
p(c_{i}\neq c_{l} \quad \hspace{-2mm}&\text{for all}\quad \hspace{-2mm} i\neq l|\boldsymbol{\text{c}}_{-i},x_{i})\propto\frac{\alpha}{N-1+\alpha}\int L(x_{i}|\boldsymbol{\theta})dG_{0}(\boldsymbol{\theta})
\label{cond_post_dirichlet2}
\end{aligned}
\end{equation}
Finally, it should be noted that the Dirichlet process mixture model can be written as follows
\begin{equation}
\begin{aligned}
x_{i}|\theta_{i}&\sim f(x_{i}|\theta_{i})\\
\theta_{i}&\sim G\\
G&\sim \text{DP}(G_{0},\alpha)
\end{aligned}
\label{DP}
\end{equation}
where $G$ is drawn from the Dirichlet process with base measure $G_{0}$.

\section{Proposed Method}
In this work we propose to extend the model developed in \cite{Punskaya02}, by assigning a class variable $c_{i}$ for the parameters $\phi_{i}$ in each segment (for $i=0,...,K$) thereby capturing the dependencies between segment parameters for improved change point estimation. This performance improvement  is achieved by concatenating  data points of segments with the same class labels; thereby providing more degrees of freedom when assessing if a change point exists. In the sections that follow, we will provide a detailed description of the proposed method. 

\subsection{Bayesian Model}
In particular, we modify the set of distinct segment  parameters, $\boldsymbol{\Phi}=[\boldsymbol{\phi}_{0},...,\boldsymbol{\phi}_{K}]$,  by introducing the class variable $c_{i}$, such that,  $\boldsymbol{\Phi}=[\boldsymbol{\phi}_{c_{0}},...,\boldsymbol{\phi}_{c_{K}}]$; where the parameters in each segment are effectively then being drawn from a set of class parameters, $\boldsymbol{\Phi}^{c}=[\boldsymbol{\phi}^{c}_{1},...,\boldsymbol{\phi}^{c}_{V}]$, where $V\leq K+1$. That is, each segment parameter can be formulated as a multivariate Gaussian mixture model (the Gaussian assumption enables tractable posterior distributions) of the class parameters $\boldsymbol{\phi}^{c}_{v}$
\begin{equation}
p(\boldsymbol{\Phi}|\boldsymbol{\Phi}^{c},\boldsymbol{\Sigma}^{c},\boldsymbol{\pi})=\sum_{v=1}^{V}\pi_{v}\mathcal{MN}(\boldsymbol{\Phi}|\boldsymbol{\phi}^{c}_{v},\Sigma^{c}_{v})
\label{proposed_mix}
\end{equation}
where $\boldsymbol{\Sigma}^{c}=[\Sigma^{c}_{1},\dots,\Sigma^{c}_{V}]$ and $\Sigma^{c}_{v}\in \mathbb{R}^{D\times D}$.
Accordingly, we present a change point estimation model that incorporates the class variable $c_{i}$ and a Dirichlet process prior on the likelihood on the class probabilities, that is 
\begin{equation}
\begin{aligned}
\boldsymbol{\text{x}}_{\tau_{i}+1:\tau_{i+1}}|\boldsymbol{\phi}^{c}_{i},\sigma^{2}_{i}&\sim f_{j}(\boldsymbol{\text{x}}_{\tau_{i}+1:\tau_{i+1}}|\boldsymbol{\phi}^{c}_{i},\sigma^{2}_{i})\\ 
\sigma^{2}_{i}&\sim G_{\sigma^{2}}\\ 
\boldsymbol{\phi}_{i}|\boldsymbol{\phi}^{c}_{i},\Sigma^{c}_{i}&\sim \mathcal{MN}(\boldsymbol{\phi}_{i}|\boldsymbol{\phi}^{c}_{i},\Sigma^{c}_{i})\\
(\boldsymbol{\phi}^{c}_{i},\Sigma^{c}_{i})&\sim G \\
G&\sim \text{DP}(G_{0},\alpha) \\
\boldsymbol{\text{x}}_{\tau_{i}+1:\tau_{i+1}}|\boldsymbol{\tau}_{K},\boldsymbol{\phi}_{i}&\sim f_{j}(\boldsymbol{\text{x}}_{\tau_{i}+1:\tau_{i+1}}|\boldsymbol{\phi}_{i})\\
 \boldsymbol{\tau}_{K}, K &\sim \text{Bin}(\boldsymbol{\tau}_{K}, K |\lambda)
\end{aligned}
\label{gen_model_changepoint}
\end{equation}
for $i=0,...,K$. Furthermore, $\text{Bin}(.)$ corresponds to a Binomial distribution, $G_{0}$ is the joint  prior distribution of both the class parameter $\boldsymbol{\phi}^{c}_{i}$ and the variance of the class parameter $\Sigma^{c}_{i}$, the prior distribution of the variance of the data points with the same class label is given by $G_{\sigma^{2}}$ and $f_{j}(.)$ corresponds to the joint Normal distribution. 
\\
\indent The posterior distribution of the model in \eqref{gen_model_changepoint},  consists  of the following parameters, $\{\boldsymbol{\tau}_{K},K,\boldsymbol{\text{c}}_{K},\hat{\boldsymbol{\Phi}^{c}},\hat{\boldsymbol{\Sigma}}^{c},\boldsymbol{\sigma}^{2}\}$, where  $\boldsymbol{\sigma}^{2}=[\sigma_{1}^{2},\dots,\sigma_{V}^{2}]$.  Inference of the parameters is carried out by using  a  Metropolis-Hastings-within-Gibbs sampling scheme. The Gibbs moves are performed on each parameter in the set, $\{\boldsymbol{\text{c}}_{K},\hat{\boldsymbol{\Phi}^{c}},\hat{\boldsymbol{\Sigma}}^{c},\boldsymbol{\sigma}^{2}\}$, while a variation of the Metropolis-Hastings algorithm is used to obtain samples for the parameters $\{\boldsymbol{\tau}_{K},K\}$. The marginal posterior distribution of parameters $\{\boldsymbol{\text{c}}_{K},\hat{\boldsymbol{\Phi}^{c}},\hat{\boldsymbol{\Sigma}}^{c},\boldsymbol{\sigma}^{2}\}$ are given by the following; namely, the marginal posterior distribution of the class parameters $\boldsymbol{\phi}^{c}_{v}$, assuming the conjugate  prior distribution, $p(\boldsymbol{\phi}^{c}_{v}|\boldsymbol{\lambda}_{\phi},\delta)\sim \mathcal{N}(\boldsymbol{\lambda}_{\phi},\delta\sigma^{2}_{v}\text{I}_{D})$, where $\boldsymbol{\lambda}_{\phi}\in \mathbb{R}^{D}$ and $\text{I}_{D})$ is an identity matrix of dimension $D$, is given by
\begin{equation}
p(\boldsymbol{\phi}^{c}_{v}|\boldsymbol{\text{c}}_{K},\boldsymbol{\tau}_{K},K,\Sigma^{c}_{v},\boldsymbol{\text{x}})\sim\mathcal{MN}(\boldsymbol{\mu}^{\phi}_{v},\Sigma^{\phi}_{v})\quad v=1,...,V
\label{cond_post_param}
\end{equation}
where $\boldsymbol{\mu}^{\phi}_{v}=\Sigma_{\phi}^{-1}\left(n_{v}\bar{\boldsymbol{\phi}}^{c}_{v}(\Sigma^{c})^{-1}+\boldsymbol{\lambda}_{\phi}^{T}\delta^{-1}\sigma^{-2}_{v}\text{I}_{D}\right)$ and $\Sigma^{\phi}_{v}=(n_{v}(\Sigma^{c})^{-1}+\delta^{-1}\sigma^{-2}_{v}\text{I}_{D})^{-1}$ for $\bar{\boldsymbol{\phi}}^{c}_{v}=\frac{1}{n_{v}}\sum_{i:c_{i}=v}\boldsymbol{\phi}_{i}^{T}$ where $n_{v}$ corresponds to the number of segment parameters $\boldsymbol{\phi}_{i}$ assigned to class $v$. The marginal posterior distribution (Inverse Wishart distributed) of the class covariance matrix $\Sigma^{c}_{v}$, given the conjugate prior distribution,  $p(\Sigma^{c}_{v}|\beta,\omega)\sim \mathcal{IW}(\beta,\boldsymbol{\Omega})$ where $\beta \in\mathbb{R}$ and $\boldsymbol{\Omega} \in\mathbb{R}^{D\times D}$, is shown by the following 
\begin{equation}
p(\Sigma^{c}_{v}|\boldsymbol{\text{c}}_{K},\boldsymbol{\tau}_{K},K,\boldsymbol{\phi}^{c}_{v},\boldsymbol{\text{x}})\sim \mathcal{IW}(\alpha^{\phi}_{v},\boldsymbol{B}^{\phi}_{v})\quad v=1,...,V
\label{cond_post_covariance}
\end{equation}
where $\alpha^{\phi}_{v}=n_{v}+\beta$ and $\boldsymbol{B}^{\phi}_{v}=\beta\boldsymbol{\Omega} + \sum_{i:c_{i}=v}(\boldsymbol{\phi}_{i}-\boldsymbol{\phi}^{c}_{i})(\boldsymbol{\phi}_{i}-\boldsymbol{\phi}^{c}_{i})^{T}$. Furthermore, the  posterior distribution of the variance $\sigma^{2}_{v}$ for  the data points from segments with the same class label $v$, along with the inverse Gamma prior distribution, $p(\sigma^{2}_{v}|\nu,\gamma)\sim \mathcal{IG}(\nu,\gamma)$, is given by (for $v=1,...,V$)
\begin{equation}
p(\sigma^{2}_{v}|\boldsymbol{\text{c}}_{K},\boldsymbol{\tau}_{K},K,\boldsymbol{\text{x}})\sim \mathcal{IW}(\nu+d_{v},\gamma+Y^{T}_{v}\boldsymbol{\text{P}}_{v}Y_{v}) 
\label{cond_post_seg_variance}
\end{equation}
where $d_{v}$ is the number of data points with label $v$, $Y_{v}$ is the concatenated vector\footnote{Furthermore, $\boldsymbol{\text{G}}_{v}$ is concatenation of input data points such that, $Y_{v}=\boldsymbol{\text{G}}_{v}\boldsymbol{\phi}^{c}_{v}$ is satisfied. } of all data points with the same segment  label $v$, $\boldsymbol{\text{P}}_{v}=\left(\mathbf{I}_{d_{v}}-\boldsymbol{\text{G}}_{v}\boldsymbol{\text{M}}_{v}\boldsymbol{\text{G}}_{v}^{T}\right)$, with $\boldsymbol{\text{M}}_{v}=(\boldsymbol{\text{G}}_{v}^{T}\boldsymbol{\text{G}}_{v}+\delta^{-1}\text{I}_{D})^{-\frac{1}{2}}$. Finally, the marginal posterior distribution for the class labels $c_{i}$ are given by: $p(c_{i}=v|\boldsymbol{\text{c}}_{-i},\boldsymbol{\phi}_{i},\boldsymbol{\phi}^{c}_{v},\Sigma_{v}^{c})$ for $n_{-i,v}>0$ (shown in \eqref{cond_post_dirichlet1}) with  likelihood $L(\boldsymbol{\phi}_{i}|\boldsymbol{\phi}^{c}_{v},\Sigma_{v}^{c})\sim \mathcal{MN}(\boldsymbol{\phi}_{i}|\boldsymbol{\phi}^{c}_{v},\Sigma_{v}^{c})$; while for the posterior probability for a new class $p(c_{i}\neq c_{l} \quad \hspace{-2mm}\text{for all}\quad \hspace{-2mm} i\neq l|\boldsymbol{\text{c}}_{-i},\boldsymbol{\phi}_{i})$ is given by \eqref{cond_post_dirichlet2} (see [] for more details). 
\\
\indent The conditional  posterior distribution of the parameters $\{\boldsymbol{\tau}_{K},K\}$ can be obtained by first considering the following marginal posterior distribution, $p(\boldsymbol{\tau}_{K},K|\lambda,\boldsymbol{\text{c}}_{K},\boldsymbol{\Phi}^{c},\boldsymbol{\sigma}^{2},\boldsymbol{\text{x}})$ where $\lambda$ is an hyperparameter of the following prior distribution,  $p(\boldsymbol{\tau}_{K},K|\lambda)=\lambda^{K}(1-\lambda)^{T-K-1}$.  Having selected the appropriate conjugate priors, we can integrate out the nuisance parameters $\{\boldsymbol{\Phi}^{c},\boldsymbol{\sigma}^{2},\lambda\}$, thereby significantly reducing the number of parameters required to specify the posterior distribution for  $\{\boldsymbol{\tau}_{K},K\}$. To this end, we first obtain the following posterior distribution (that incorporates the prior distributions of the nuisance parameters)
\begin{equation}
\begin{aligned}
&p(\boldsymbol{\Phi}^{c},\boldsymbol{\sigma}^{2},\boldsymbol{\tau}_{K},K,\lambda|\boldsymbol{\text{c}}_{K},\mathbf{x})\propto p(\mathbf{x}|\boldsymbol{\Phi}^{c},\boldsymbol{\sigma}^{2},K,\boldsymbol{\tau}_{K},\boldsymbol{\text{c}}_{K})\times p(K,\boldsymbol{\tau}|\lambda)p(\lambda)\prod_{v=1}^{V}p(\boldsymbol{\phi}^{c}_{v}|\boldsymbol{\lambda}_{\phi},\delta)p(\sigma^{2}_{v}|\nu,\gamma)
\label{posterior_proposed}
\end{aligned}
\end{equation}
where  $p(\lambda)$ has uniform probability over the interval $[0,1]$ and  the likelihood function is given by 
\begin{equation*}
p(\mathbf{x}|\boldsymbol{\Phi}^{c},\boldsymbol{\sigma}^{2},K,\boldsymbol{\tau}_{K},\boldsymbol{\text{c}}_{K})=\prod_{v=1}^{V}\prod_{i:c_{i}=v}p(\mathbf{x}_{\tau_{i}+1:\tau_{i+1}}|\boldsymbol{\phi}^{c}_{v},\sigma_{v}^{2})
\label{likelihood_proposed}
\end{equation*}
where by combining data points from the same segment class label $v$, we can potentially obtain more accurate parameter estimation owing to the increased number of number available for estimating $\{\boldsymbol{\tau}_{K},K\}$. Integration of \eqref{posterior_proposed} with respect to the parameters $\{\boldsymbol{\phi}^{c}_{v},\sigma_{v}^{2},\lambda\}$ results in the following expression for the conditional posterior distribution of the parameters $\{\boldsymbol{\tau}_{K},K\}$
\begin{equation}
\begin{aligned}
&p(\boldsymbol{\tau}_{K},K|\boldsymbol{\text{c}}_{K},\mathbf{x})\propto \prod_{v=1}^{V}\frac{2^{\frac{\nu}{2}}}{\Gamma(\frac{\nu}{2})}\Gamma(K+1)\Gamma(N-K+1) \left(\frac{\gamma}{2}\right)^{\frac{\nu}{2}}  \times\Gamma\left(\frac{d_{v}+\nu}{2}\right)\pi^{-\frac{d_{v}}{2}}\left[\gamma+Y^{T}_{v}\boldsymbol{\text{P}}_{v}Y_{v}\right]^{-\frac{d_{v}+\nu}{2}}|\boldsymbol{\text{M}}_{v}|^{-\frac{1}{2}}
\end{aligned}
\label{full_posterior_proposed}
\end{equation}
 Finally, we note that there are some challenges from drawing samples from \eqref{full_posterior_proposed} due to the dependence on $\boldsymbol{\text{c}}_{K}$ that we have addressed in the next section.
\subsection{Gibbs Sampling}
\begin{algorithm}[b!]
\begin{algorithmic} 
\REQUIRE 
\STATE - Select: Set input parameters (discussed in Section IV).
\STATE - Initialize: $\{\boldsymbol{\tau}_{K},K,\boldsymbol{\text{c}}_{K},\hat{\boldsymbol{\Phi}^{c}},\hat{\boldsymbol{\Sigma}}^{c},\boldsymbol{\sigma}^{2},V\}$
\STATE - Set: $N_{iter}$ and $N^{c}_{iter}$.
\vspace{1mm}
\FOR{$loop_{main}=\{1,\dots,N_{iter}\}$} 
\FOR{$loop_{c}=\{1,\dots,N^{c}_{iter}\}$} 
\STATE -  Sample $p(\boldsymbol{\phi}^{c}_{v}|\boldsymbol{\text{c}}_{K},\boldsymbol{\tau}_{K},K,\Sigma^{c}_{v},\boldsymbol{\text{x}})$ for $ v=1,..V$, shown in \eqref{cond_post_param}.
\STATE -  Sample $p(\Sigma^{c}_{v}|\boldsymbol{\text{c}}_{K},\boldsymbol{\tau}_{K},K,\boldsymbol{\phi}^{c}_{v},\boldsymbol{\text{x}}) $ for $v=1,...,V$, shown in \eqref{cond_post_covariance}.
\STATE  - Sample $p(\sigma^{2}_{v}|\boldsymbol{\text{c}}_{K},\boldsymbol{\tau}_{K},K,\boldsymbol{\text{x}})$ for $v=1,...V$, shown in \eqref{cond_post_seg_variance}.
\STATE - Sample class variable $c_{i}$ for $i=0,...,K$, using \eqref{cond_post_dirichlet1} and \eqref{cond_post_dirichlet2}.
\ENDFOR
\STATE - Sample $p(\boldsymbol{\tau}_{K},K|\boldsymbol{\text{c}}_{K},\mathbf{x})$, refer to Section III.B.
\ENDFOR
\end{algorithmic}
\caption{}\label{alg:Algo1}
\end{algorithm}
A summary of the Gibbs sampling scheme for drawing samples for the parameters $\{\boldsymbol{\tau}_{K},K,\boldsymbol{\text{c}}_{K},\hat{\boldsymbol{\Phi}^{c}},\hat{\boldsymbol{\Sigma}}^{c},\boldsymbol{\sigma}^{2}\}$, is provided in Algorithm \ref{alg:Algo1}. Observe that the parameters $\{\boldsymbol{\tau}_{K},K\}$, are dependent on the segment class variables $\{\boldsymbol{\text{c}}_{K}\}$, and in turn, the segment class variables are dependent on the parameters $\{\boldsymbol{\tau}_{K},K,\hat{\boldsymbol{\Phi}^{c}},\hat{\boldsymbol{\Sigma}}^{c},\boldsymbol{\sigma}^{2}\}$. Furthermore, the marginal posterior distributions for both $\{\boldsymbol{\tau}_{K},K\}$ and $\{\boldsymbol{\text{c}}_{K}\}$ are intractable and therefore require sampling schemes; in particular, a nested Gibbs sampling scheme was used in order to draw samples for the class variables $\{\boldsymbol{\text{c}}_{K}\}$, due to the dependence on the  parameters $\{\hat{\boldsymbol{\Phi}^{c}},\hat{\boldsymbol{\Sigma}}^{c},\boldsymbol{\sigma}^{2}\}$ (as shown in Algorithm \ref{alg:Algo1}). While, A modification of the Metropolis-Hastings algorithm (having integrated out the nuisance parameters) outlined in \cite{Punskaya02} was used in order to draw samples from the conditional posterior distribution,  $p(\boldsymbol{\tau}_{K},K|\boldsymbol{\text{c}}_{K},\mathbf{x})$; in particular, a variation was developed that incorporates the   segment labels $\boldsymbol{\text{c}}_{K}$. 
\\
\indent Given the $j^{th}$ samples, $\{\boldsymbol{\tau}_{K},K\}_{j}$,  we first select with a certain probability, one the following:
\begin{itemize}
  \item  $K\rightarrow K+1$: create a new change point (birth), with probability, $b$
  \item  $K\rightarrow K-1$:  remove an existing change (death) with probability, $d$
 \item $K\rightarrow K$: update of change point positions with probability, $u$
\end{itemize}
where   $b=d=u$ for $0<K<K_{max}$, and $b+d+u=1$ for $0\leq K\leq K_{max}$. Furthermore, for $K=0$, $d=0$ and $b=u$, while for $K=K_{max}$, $b=0$ and $d=u$.
\\
\indent  A birth  move consists of proposing a new transition time $\tau_{prop}$, with the following proposal distribution, $q(\boldsymbol{\tau}_{K+1}|\boldsymbol{\tau}_{K})=q(\tau_{prop}|\boldsymbol{\tau}_{K})= \{\frac{1}{N-K-2}\quad \text{for} \quad \tau_{prop} \in S_{prop}\}$, where $\boldsymbol{\tau}_{K+1}$ corresponds to the set of  change points that includes both $\tau_{prop}$ and $\boldsymbol{\tau}_{K}$, while $S_{prop}$ corresponds to the set of time indices $[2,N-1]$ excluding the time points $\boldsymbol{\tau}_{K}$. Conversely, the proposal distribution for removing  $\tau_{prop}$ from $\boldsymbol{\tau}_{K+1}$, is given by  $q(\boldsymbol{\tau}_{K}|\boldsymbol{\tau}_{K+1})=\{\frac{1}{K+1} \quad \text{for} \quad \tau_{prop}\in \boldsymbol{\tau}_{K+1}\}$. Accordingly, the proposed transition time $\tau_{prop}$ is accepted with the following probability, $\alpha_{birth}=\text{min}\{1,r_{birth}\}$,
\begin{equation*}
r_{birth}=\frac{p(\boldsymbol{\tau}_{K+1},K+1|\boldsymbol{\text{c}}_{K+1},\mathbf{x})}{p(\boldsymbol{\tau}_{K},K|\boldsymbol{\text{c}}_{K},\mathbf{x})}\frac{q(\boldsymbol{\tau}_{K}|\boldsymbol{\tau}_{K+1})q(K|K+1)}{q(\boldsymbol{\tau}_{K+1}|\boldsymbol{\tau}_{K})q(K+1|K)}
\label{accept_probab}
\end{equation*}
\begin{figure}[t!]
\begin{center}
\includegraphics[width=0.7\columnwidth]{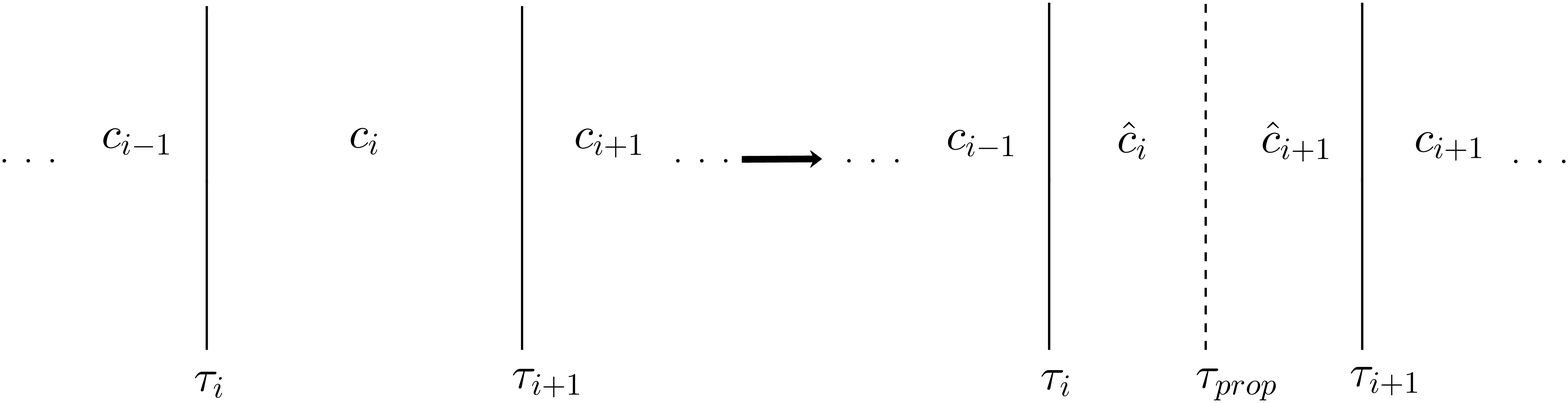}
\includegraphics[width=0.7\columnwidth]{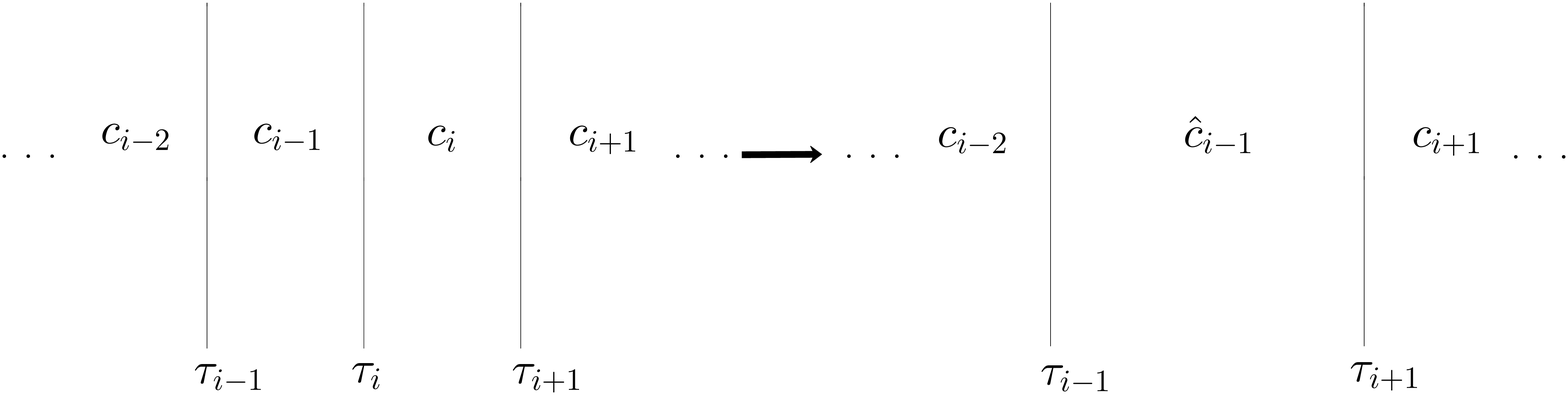}
\caption{Detecting changes in power data with repeating patterns, using the proposed method (upper panel) and MCMC method in \cite{Punskaya02} (lower panel). 
 }\label{fig:power}
\end{center}
\end{figure}
with $q(K+1|K)=b$ and $q(K|K+1)=d$, corresponding to the proposal distributions for the unit increment and decrement (respectively) of the parameter $K$. It should be noted that in order to determine the  acceptance ratio $r_{birth}$, we need to evaluate $p(\boldsymbol{\tau}_{K+1},K+1|\boldsymbol{\text{c}}_{K+1},\mathbf{x})$, where  there is now a  dependence on  $\boldsymbol{\text{c}}_{K+1}$. This dependence arises due to the proposed transition time  $\tau_{prop}$, splitting the segment between the   time indices $\{\tau_{i},\tau_{i+1}\}$ into $\{\tau_{i},\tau_{prop},\tau_{i+1}\}$, as well as, splitting  the  segment class variable $\{c_{i}\}$, into two new class variables  $\{\hat{c_{i}},\hat{c}_{i+1}\}$.  As we have not yet inferred the new class variables from the conditional  class posterior distributions,  we assume that the two classes $\{\hat{c_{i}},\hat{c}_{i+1}\}$ are distinct (that is, $\hat{c}_{j}\neq c_{k}$ for all $j\neq k$ and $j=1,2$) and thus independent  from all other segments, to circumvent  the lack of information we have for assignment to an existing class (please refer to Figure ). 
\\
\indent The death move proposes to remove a transition time $\tau_{prop}$, by choosing with uniform probability from the set $\boldsymbol{\tau}_{K}$; where the removal of $\tau_{prop}$ is accepted with probability   $\alpha_{death}=\text{min}\{1,r_{birth}^{-1}\}$. As in the previous case (birth move), we need to determine $r_{birth}$, however, now we need to evaluate $p(\boldsymbol{\tau}_{K-1},K-1|\boldsymbol{\text{c}}_{K-1},\mathbf{x})$.  That is, the segments between the transition times, $\{\tau_{i},\tau_{prop}\}$ and $\{\tau_{prop}, \tau_{i+2}\}$ where $\tau_{prop}=\tau_{i+1}$, are combined into one segment $\{\tau_{i},\tau_{i+2}\}$, along with the  segment class variables $\{c_{i},c_{i+1}\}$ being combined into one segment with a new class variable $\{\hat{c}_{i}\}$. Using the argument utilised for the birth  of a change point we assign a distinct value to the new class variable, that is, $\hat{c}_{i}\neq c_{j}$ for all $j\neq i$. 
\\
\indent The update of the transitions times is carried by first removing the $j^{th}$ transition time index  $\tau_{j}$ from $\boldsymbol{\tau}_{K}$ and proposing a new change point at some new location, for all $ j=\{1,\dots,K\}$. That is, the death move is first applied followed by  a birth move for all transition times in $\boldsymbol{\tau}_{K}$.

\bibliographystyle{IEEEtran}
\bibliography{sigproc}
\end{document}